\documentclass[12pt]{article}
\usepackage{pst-all}
\usepackage{amsfonts}
\usepackage{latexsym}
\usepackage{amsmath,amssymb}
\usepackage{verbatim}
\usepackage{datetime}

\newcommand{\bea}{\begin{eqnarray}}
\newcommand{\eea}{\end{eqnarray}}
\newcommand{\be}{\begin{equation}}
\newcommand{\ee}{\end{equation}}
\newcommand{\ba}{\begin{array}}
\newcommand{\ea}{\end{array}}
\newcommand{\lb}{\left(}
\newcommand{\rb}{\right)}

\def\nn{\nonumber}
\def\half{{1\over2}}
\def\p{\partial}
\def\eb{\lambda}
\def\g{\gamma}
\def\bg{\bar{\gamma}}

\def\bx{\bar{x}}

\def\bz{\bar{z}}
\def\bw{\bar{w}}

\def\b#1{\bar{#1}}

\def\bp{\bar{\p}}
\def\tg{\hat{\g}}
\def\bq{\bar{q}}

\numberwithin{equation}{section}
\begin{document}
\begin{center}
{ \LARGE {{The Spectrum of Strings on Warped AdS$_3 \times$S$^3$}}}

\vspace{0.8cm}

\vspace{1cm}

Tatsuo Azeyanagi, Diego M.~Hofman, Wei Song and Andrew Strominger
\vspace{0.5cm}

{\it  Center for the Fundamental Laws of Nature, Harvard University,\\
Cambridge, MA 02138, USA}

\vspace{0.5cm}

\vspace{1.0cm}

\end{center}

\begin{abstract}
String theory on NS-NS AdS$_3\times$S$^3$ admits an exactly marginal deformation which  breaks the $\overline{ S L(2,\mathbb{R})}_R\times SL(2,\mathbb{R})_L$ isometry of  AdS$_3$ down to $\overline {SL(2,\mathbb{R})}_R \times  U(1)_L $. The holographic dual is an exotic and only partially understood type of  two-dimensional CFT with a reduced unbroken global conformal symmetry group. In this paper we study the deformed theory on the string worldsheet.  It is found to be related by a spectral flow which is nonlocal in spacetime to the undeformed worldsheet theory.  An exact formula for the spectrum of massive strings is presented.  
\noindent
\end{abstract}
\thispagestyle{empty}

\pagebreak
\setcounter{tocdepth}{3}

\tableofcontents
\section{Introduction}

Two dimensional conformal field theories with an $ SL(2,\mathbb{R})_L\times \overline{ S L(2,\mathbb{R})}_R$ global symmetry are of central interest for a  wide variety of applications in physics and mathematics. Interest in a possible new class of exotic 2D conformal field theories with a smaller group of $ U(1)_L\times \overline{ S L(2,\mathbb{R})}_R$ global symmetries, including right but not left dilations,  has recently arisen in several disparate contexts: topologically massive gravity \cite{Anninos:2008fx}, Kerr/CFT \cite{Guica:2008mu}, nonlocal dipole theories \cite{Bergman:2000cw, Dasgupta:2000ry,Bergman:2001rw,
ElShowk:2011cm,Song:2011sr} and AdS/CMT with Schr\"odinger spacetimes \cite{Son:2008ye,Balasubramanian:2008dm,Hartnoll:2009sz}.   Quantum theories of gravity on warped AdS$_3$ which have $ U(1)_L\times \overline{ S L(2,\mathbb{R})}_R$ isometries potentially  provide holographic duals.

Understanding these exotic 2D theories and their holographic duals is a very challenging, but possibly soluble, problem. At present, there is no clearly understood example. Here  we consider what may be the most tractable nontrivial example. It is
a deformation of string theory on $AdS_3\times S^3\times{M}^4$ supported with NS-NS flux, where ${M}^4\,(T^4 \,\hbox{or} \,K3)$ is a four dimensional compact manifold. The deformation is constructed via the solution-generating so-called TsT transformation studied in
\cite{Lunin:2005jy,Maldacena:2008wh}, which warps both the $AdS_3$ and the $S^3$ and  breaks the isometries as\footnote{Throughout this paper we use a bar to denote right-moving quantities in both spacetime and the worldsheet.}
\bea SL(2,\mathbb{R})_L\times SU(2)_L& \times &\overline{ S L(2,\mathbb{R})}_R\times \overline {SU(2)}_R  \cr  \to U(1)_L\times SU(2)_L &\times &\overline {SL(2,\mathbb{R})}_R\times \overline {U(1)}_R\, . \eea
The unbroken $U(1)_L$ is noncompact.
We consider mainly the case where it is null and corresponds to left translations.\footnote{{The time-like case is discussed in a different setup \cite{Detournay:2010rh}.} } The word deformation must be used with some care in this case, because the deformation parameter can be set to unity  by a coordinate transformation. It is therefore not a priori self-evident in what sense the deformed and undeformed theory are close.

The spectrum of string theory in the unwarped theory was described in \cite{Maldacena:2000hw}, building on the earlier work \cite{Balog:1988jb, Bars:1990rb, Evans:1998qu, Bars:1995cn} . The main result  of the present paper is a description of the warped spectrum. This is possible because of the beautiful work of \cite{Frolov:2005dj,Alday:2005ww} mapping the deformed worldsheet for a general TsT-deformed geometry to the undeformed worldsheet with twisted boundary conditions which are nonlocal in spacetime.  The deformed massive string spectrum is then related to the undeformed spectrum by a spectral flow whose parameters depend on the $U(1)_L\times \overline {U(1)}_R$ charges of the string mode.

The massless spectrum - which includes boundary gravitons and gluons of various kinds - is more subtle.
The worldsheet analysis constrains but does not fully determine this spectrum.  A full determination requires imposition of consistent asymptotic spacetime boundary conditions.
In both supergravity and string theory it is certainly possible to construct linearized one-particle excitations which violate the boundary conditions, so the mere existence of a linearized solution to the equation of motion does not imply its physical inclusion. Moreover,  examples are known 
\cite{Klebanov:1999tb} in string theory with more than one consistent set of boundary conditions for the same worldsheet theory, so there can not in general be a unique answer to which worldsheet modes should be included. In this paper we will stick to the pure worldsheet analysis and will not address spacetime boundary conditions. These boundary conditions and the associated holographic renormalization have been recently found to have an extremely interesting and subtle structure \cite{Guica:2011ia,balt}. We expect the worldsheet results herein to usefully inform and constrain future efforts to completely  determine the consistent boundary conditions and associated spectra. For the boundary modes, we herein construct the vertex operators and describe how their asymptotic falloffs and associated symmetry generators change under the deformation.

Prior to deformation, the free multi-string Hilbert space forms a representation of  both a left and a right spacetime Virasoro-SU(2)-Kac-Moody algebra.  We show that the deformation manifestly preserves  the action of a right Virasoro-U(1)-Kac-Moody, as well as a global left-moving
$ U(1)_L\times SU(2)_L  $. This strongly suggests that the spacetime theory is a conformal field theory with (at least) one Virasoro action.  The fate of the other Virasoro-Kac-Moody generators depends on spacetime boundary conditions. We also investigate the surprising possibility of ``crossover"  Kac-Moody symmetries, which transform under the right-moving  Virasoro but have zero modes which are left isometries. In particular we construct worldsheet weight $(1,1)$ vertex operators that form a tower of $U(1)_C\times SU(2)_C$ right-moving Kac-Moody boundary photons and gluons. We show that the zero mode is the unbroken $U(1)_L\times SU(2)_L$, and compute the level of the associated Kac-Moody algebra. Again whether or not the crossover modes should be physically included is a spacetime question outside the scope of the present paper.

While the crossover phenomenon sounds bizarre at first, it has in fact been previously encountered in a variety of contexts. For example \cite{Strominger:1998yg,Hartman:2008dq} considered circle compactification of
$AdS_3$ to $AdS_2$ which preserved an $\overline{ S L(2,\mathbb{R})}_R$ isometry algebra along with its  right-moving  Virasoro enhancement, which acts as local conformal transformations in $AdS_2$. The circle isometry is the unbroken $U(1)_L$ subgroup of the $S L(2,\mathbb{R})_L$  $AdS_3$ isometry group. After compactification, it is the global gauge transformation of the Kaluza-Klein $U(1)_L$ gauge field. This is promoted to a local gauge symmetry in $AdS_2$ which is a left $\to$ right crossover $U(1)_C$ Kac-Moody transforming under the right Virasoro. Furthermore, this symmetry structure may ultimately be responsible \cite{STD} for the Cardy-like expressions for the entropy of various types of warped black holes.

An interesting feature of these theories is that they lie at the crossroads of several current avenues of research.  They have been investigated from a variety of complimentary viewpoints, other than the worldsheet perspective developed herein. From the point of view of the boundary conformal field theory Sym$^{Q^2}(M^4)$, the deformation is generated by a certain dimension $(1,2)$ operator constructed from a chiral primary \cite{ElShowk:2011cm}. Holographic renormalization in Schr\"odinger spacetimes (of which this is a special case) has been studied in supergravity \cite{Guica:2010sw,Guica:2011ia,balt}: our results agree with these in the regime of overlapping validity.  2D theories with $\overline{S L(2,\mathbb{R})}_R\times U(1)_L$ global symmetries have been analyzed using purely field theoretic methods in \cite{Hofman:2011zj}.  The asymptotic symmetry group can likely be analyzed with supergravity methods similar to the ones employed in \cite{ElShowk:2011cm,Song:2011sr}.  The TsT
transformation yields theories similar to  the dipole-deformed gauge theories \cite{Bergman:2000cw,Dasgupta:2000ry,Bergman:2001rw}, which are TsT transformations of RR backgrounds. These are fascinating non-local theories  similar in spirit  but perhaps simpler than non-commutative gauge theories. It would certainly be interesting to see how the dipole star product enters the game.  The existence of all these different approaches suggests that we will be able to learn a lot.

This paper is organized as follows. In section 2 we briefly review the original work of \cite{Giveon:1998ns, Kutasov:1999xu}, on which we heavily rely throughout, on worldsheet string theory and AdS/CFT on NS-NS $AdS_3\times S^3$.
In section 3 we present the null warped solution obtained via a TsT transformation, and derive the modified conformal weight of an operator dual to a charged massive scalar. In section 4 we show that the string worldsheet can be transformed by a field redefinition 
back to unwarped $AdS_3\times S^3$ at the price of twisted boundary conditions. The twists  are reinterpreted as spectral flow and an exact formula for the modified spectrum of massive string states thereby derived. In section 5 we demonstrate agreement at low curvatures between the string and supergravity formulae for the spectrum. In section 6 we describe the fate of the various boundary gluons and gravitons related to Virasoro Kac-Moody symmetries in the undeformed theory, as well as the crossover modes. In section 7 we briefly discuss the finite temperature properties of the theory.

\section{Review of string theory on $AdS_3\times S^3$ with NS-NS field}
We start with the solution of string theory $AdS_3\times S^3\times{M}^4$ supported with NS-NS flux, where ${M}^4\,(T^4 \,\hbox{or} \,K3)$ is a four dimensional compact manifold. The six dimensional part of the solution is
\bea ds^2&=&Q(d\rho^2+e^{2\rho}d\gamma d\bar{\gamma}+d\Omega_3^2),\nn\\
B&=&-{Q\over4} (\cos\theta d\phi \wedge d\psi+{2e^{2\rho}}d\gamma \wedge d\bar{\gamma} ),\label{bg}
\eea
where $d\Omega_3$ is the line element on a unit three sphere. In this paper we work in Lorentzian signature, therefore $\gamma$ and $\bar{\gamma}$ should be thought as independent left and right-moving coordinates, instead of being complex conjugate to each other.\footnote{As our deformation breaks Lorentz invariance, it is awkward to work in Euclidean space. We nevertheless use an overbar in Lorentz signature to distinguish right-moving quantities in conformity with the notation of  
\cite{Giveon:1998ns, Kutasov:1999xu}.} This solution can also be obtained as the near horizon geometry of the NS1-NS5 configuration \cite{Maldacena:1998bw}, with $Q$ NS1 branes and $Q$ NS5 branes.
The background preserves $SL(2,\mathbb{R})_L\times \overline{SL(2,\mathbb{R})}_R\times SU(2)_L\times \overline{SU(2)}_R$ isometry.

The worldsheet Lagrangian is
\bea \mathcal{L}&=&-{1\over4\pi}\sqrt{-h}(h^{ab}g_{\mu\nu}\p_a X^\mu\p_b X^\nu+\epsilon^{ab}B_{\mu\nu}\p_aX^\mu \p_bX^\nu)\nn\\&=&{Q\over2\pi}\lb e^{2\rho}\p \bg\bar{\p}{\gamma}+\p \rho \bar{\p}\rho+{1\over4}(\bar{\p}\psi+2\cos\theta\bar{\p}\phi){\p}\psi+\cdots\rb, \label{ac}\eea
where $z=\tau+\sigma$ ($\bar z =\tau-\sigma$) is a left (right) - moving null coordinate and $\cdots$ denotes terms invariant under shift of $\gamma$ and $\psi$.The $AdS_3$ part of the action can be rewritten in terms of the auxiliary fields
\bea
S&=&{1\over2\pi}\int d^2z \left(Q \p\rho\bar{\p}\rho+\beta\bar{\p}\gamma+\bar{\beta}\p\bar{\gamma}-{\beta\bar{\beta}\over Q}e^{-2\rho}\right).
\eea
At the boundary of $AdS_3,\,\rho\rightarrow\infty$, the interaction term $\beta\bar{\beta}e^{-2\rho}$ is suppressed, and the $AdS_3$ part of the theory can be approximated by the  left-moving  free fields $(\beta(z),\gamma(z))$, the right-moving free fields $(\bar{\beta}(\bar z),\bar{\gamma}(\bar z))$, and the non-chiral free fields $\rho(z, \bar z)$.

\subsection{Conserved currents }
At the level of the classical sigma model, the $SL(2,\mathbb{R})$ currents are
\bea
j^-&=&-Qe^{2\rho}\p \bg,\quad j^3=-{Q}(e^{2\rho}\g\p\bg-\p \rho),\quad j^+=-{Q}(e^{2\rho}\g^2\p\bg-2\g\p\rho-\p\g), \nn\\
\bar{j}^-&=& -Qe^{2\rho}\bar{\p} \g,\quad \bar{j}^3=-{Q}(e^{2\rho}\bg\bar{\p}\g-\bar{\p} \rho),\quad \bar{j}^+=-{Q}(e^{2\rho}\bg^2\bar{\p}\g-2\bg\bar{\p}\rho-\bar{\p}\bg), \nn\\
\eea
where currents with bars are right-moving while currents without bars are left-moving.
The classical $SU(2)$ currents are
\bea
k^3&=&-i{Q\over2}(\p\phi+\cos\theta\p\psi),\quad k^\pm={Q\over2}e^{\mp i\phi}(\p\theta\pm i\sin\theta \p\psi),\\
\bar{k}^3&=&-i{Q\over2}(\bar{\p}\psi+\cos\theta\bar{\p}\phi),\quad \bar{k}^\pm={Q\over2}e^{\mp i\psi}(\bar{\p}\theta\pm i\sin\theta \bar{\p}\phi).\eea

\subsection{The spectrum}
The spectrum of the $ SL(2,\mathbb{R}) $ WZW model is given in \cite{Maldacena:2000hw} building on earlier works including  \cite{Balog:1988jb, Bars:1990rb, Evans:1998qu, Bars:1995cn}. The Hilbert space is decomposed into the representation of $SL(2,\mathbb{R})_L\times \overline{SL(2,\mathbb{R})}_R$ current algebra
\be \mathcal{H}_0=\oplus_{w=-\infty}^\infty\left[ (\int_{\half}^{Q-1\over2}dj\mathcal{D}_j^w\otimes\mathcal{D}_j^w)\oplus(\int_{\half+i\mathbb{R}}
dj\int_0^1d\alpha C_{j,\alpha}^w\otimes C_{j,\alpha}^w)\right].\ee
$\mathcal{D}_j^w$ is an irreducible representation of the $SL(2,\mathbb{R})$ generated by
the highest weight representation, while $C_{j,\alpha}^w$ is generated from the principal continuous  representation. $ w$ is the winding number. $\mathcal{D}_j^w$ and $C_{j,\alpha}^w$ are related to $\mathcal{D}_j^0$ and $C_{j,\alpha}^0$ by spectral flow.

\subsection{Primary vertex operators}
The basic building block for worldsheet vertex operators is
the boundary primary operator with conformal weight ($h,\bar{h}$)=($h,h$) under the $SL(2,\mathbb{R})_L\times \overline{SL(2,\mathbb{R})}_R$
given by \cite{Kutasov:1999xu} \be \Phi_h(x,\bx;z,\bz)={1\over\pi}\left({1\over(\gamma-x)(\bar{\gamma}-\bar{x})e^\rho+e^{-\rho}}\right)^{2h},\ee
where $(x, \bar{x})$ are the left and right null coordinates of the boundary CFT, while $\gamma,~\bar \gamma$ and $\rho$ are worldsheet operators depending on the worldsheet coordinates $(z,\bz)$. 
In momentum space, we have
\bea \Phi_{h}(p,\bx;z,\bz)&\equiv& \int {dx\over{2\pi}} e^{ipx} \Phi_{h}(x,\bar{x};z,\bar{z})\\
&=&{(-1)^{h}p^{2h-1}\over\pi\Gamma(2h)(\bg-\bx)^{2h}e^{2h\rho}}e^{ip(\gamma+{e^{-2\rho}\over \bg-\bx})}, \eea
and
\bea
\Phi_{h}(p,\bar{p};z,\bz)&\equiv&
\int {d\bar{x}\over{2\pi}} e^{i\bar{p}\bar{x}} \Phi_{h}(p,\bar{x};z,\bar{z})\\
&=&{(p\bar{p})^{h-\half}\over\pi\Gamma(2h)}e^{i(p\gamma+\bar{p}\bg)}e^{-\rho} I_{2h-1}(2e^{-\rho}\sqrt{p\bar{p}}).
\eea
Note that the modified Bessel function of the first kind $I_{2h-1}(2e^{-\rho}\sqrt{p\bar{p}})$ goes to ${(p\bar{p})^{h-\half}e^{-(2h-1)\rho}\over\Gamma(2h)}$ as $\rho\rightarrow\infty.$

\section{Supergravity analysis }
This section describes the warped supergravity solution and the deformation of its spectrum of linearized excitations. In particular we generalize to the charged case the formula of 
\cite{Guica:2010sw} for the boundary conformal weight of the operator dual to a scalar.
The solution we are interested in is a TsT transformation of $AdS_3\times S^3$.  To obtain it we compactify and then
T dualize along $\gamma$, shift $\psi\rightarrow\psi-2 {\lambda\over Q} \gamma$, T dualize along $\gamma$ again, decompactify and
then shift $\gamma\rightarrow \gamma+{\lambda \over2}\psi$. The new  gravity solution is
\bea
ds^2&=&Q\lb d\rho^2+e^{2\rho}d\gamma d\bar{\gamma}+d\Omega_3^2+{\eb} e^{2\rho}d\bar{\gamma}(d\psi+\cos\theta d\phi)\rb,
\label{TsTbg}\\
B&=&-{Q\over4} \lb \cos\theta d\phi \wedge d\psi+{2e^{2\rho}}d\gamma \wedge d\bar{\gamma} +2\eb e^{2\rho}(d\psi+\cos\theta d\phi)\wedge d\bg\rb.\nn
\eea
The new background has the reduced isometry  group $U(1)_L\times SU(2)_L\times \overline{{SL(2,\mathbb{R})}}_R\times \overline{U(1)}_R.$
Dimensional reduction on the three sphere gives a null warped $AdS_3$ in three dimensions,
with the metric \be ds^2_3=Q(d\rho^2+e^{2\rho}d\g d\bg-\lambda^2e^{4\rho}d\bg^2)\ee
and an $\overline{SU(2)}_R$ gauge field
\be A^3_R=2\lambda e^{2\rho}d\bg,~~~~~F_R^3=2 \, Q^{-\frac{1}{2}} *A_R^3,\ee
breaking $\overline{SU(2)}_R$ to $\overline{U(1)}_R$.

In the supergravity approximation, scalar waves on the deformed background (\ref{TsTbg}) can be expanded as \be\Phi=e^{i(\bar{p}\bar{\g}+p\g+{\bar{q}\over2}\psi+{q\over2}\phi)}Z(\rho)\Theta(\theta).\ee
The wave equation is separable. The radial equation and angular equation are
\bea
KZ(\rho)&=&Z''(\rho)+2Z'(\rho)-(M^2Q-4\lambda p\bar{q}+4\lambda^2 p^2)Z(\rho)-4p\bar{p}e^{-2\rho}Z(\rho),\nn\\
\\
-{K}\Theta(\theta)&=&4\Theta''(\theta)+4\cot\theta \Theta'(\theta)-\lb\bar{q}^2+q^2-2q\bar{q}\cos\theta\rb{\Theta(\theta)\over\sin^2\theta},\nn
\eea
where $M$ is the mass in six dimensions, and $K$ is the separation constant.
One finds that the radial equation is changed only by the shift \be M^2Q+K\rightarrow M^2Q+K-4\lambda p\bar{q}+4\lambda^2 p^2.\ee
The angular equation  is independent of the deformation and remains that of the round $S^3$. Therefore for highest weight representation, the quantization condition for the highest weight state is \bea K&=&|\bar{q}|(|\bar{q}|+2),\quad \hbox{for}\quad |\bar{q}|>|q|,\\
K&=&|q|(|q|+2),\quad \hbox{for}\quad |\bar{q}|<|q|
.\eea
The falloff is modified accordingly and the  $\overline{ S L(2,\mathbb{R})}_R$ weight becomes \bea \bar{h}_{sugra}&=&\half\left(1+\sqrt{1+M^2Q+K-4\lambda p\bar{q}+4\lambda^2 p^2}\right) \cr
&\underset{{\rm for}~|\bar q |>|q|}{=}&  \half\left(1+\sqrt{1+M^2Q+{2|\bar q| }+(\bar q-2\lambda p )^2} \right). \label{hlsugra}\eea

\section{String analysis}

In this section we map, following \cite{Frolov:2005dj,Alday:2005ww}, string theory on the TsT-deformed background back to string theory on $AdS_3\times S^3$ with twisted boundary conditions. The twisting involves the left momentum $p$ and the right charge $\bar q$ and so we work in terms of $(p, \bar q)$ eigenstates.  We then use this map to derive an exact formula for the deformed massive spectrum in terms of the undeformed one. Massless modes will be discussed in section 6.

\subsection{Nonlocal field redefinition
}

The string worldsheet Lagrangian on the background (\ref{TsTbg}) is, to leading order in the large-$Q$ sigma-model expansion
\bea \mathcal{L}&=&{Q\over2\pi}\lb e^{2\rho}{\p} \bg(\bar{\p}{\gamma}+\eb(\bar{\p}\psi+\cos\theta \bar{\p}\phi))+\p \rho \bar{\p}\rho+{1\over4}(\bar{\p}\psi+2\cos\theta\bar{\p}\phi){\p}\psi+\cdots\rb,\nn\\\label{TsTac}\eea
with the constraints\bea -{T_{WS}(z)\over Q}&=& e^{2\rho}\p\g\p\bg+{1\over4}(\p\psi+\cos\theta\p\phi)^2+{\eb}e^{2\rho}\p\bg(\p\psi+\cos\theta\p\phi)+\cdots,\nn\\
-{\bar T_{WS}(\bar z)\over Q}&=& e^{2\rho}\bar{\p}\g\bar{\p}\bg+{1\over4}(\bar{\p}\psi+\cos\theta\bar{\p}\phi)^2+{\eb}e^{2\rho}\bar{\p}\bg(\bar{\p}\psi+\cos\theta\bar{\p}\phi)+\cdots.\nn\\ \eea
Let us now consider the field redefinition\footnote{As is argued in \cite{Alday:2005ww}, we can make a gauge transformation to uncharge the fermions. Therefore it is sufficient to only consider the bosonic part.}\bea \label{redef}
\p\hat{\g}&=&\p\g-{\eb^2}e^{2\rho}\p\bg,\nn\\
\bar{\p}\hat{\g}&=&\bar{\p}\g+{\eb} (\bar{\p}\psi+\cos\theta\bar{\p}\phi),\nn\\
\p\hat{\psi}&=&\p\psi+2\eb e^{2\rho}\p\bg,\nn\\
\bar{\p}\hat{\psi}&=&\bar{\p}{\psi}.
\eea
Inserting this field redefinition into the equation of motion and constraints eliminates $\lambda$ and reverts
local worldsheet dynamics to that of $AdS_3\times S^3$. In particular there is a full $SL(2,\mathbb{R})_L\times SU(2)_L \times \overline{ S L(2,\mathbb{R})}_R\times \overline {SU(2)}_R$  set of locally conserved worldsheet currents.

The integrated form of the field redefinition can be written as
\bea
\hat{\gamma}&=&\gamma-\lambda^2 {\mu\over Q}+\lambda \bar{\varphi},\\
\hat{\psi}&=&\psi+2\lambda {\mu\over Q},
\label{frd}\eea
where $\mu$ is the bosonization of the current  $j^-$, namely
 \be {\p} \mu=-j^-=Qe^{2\rho} \p \bar{\g}=\beta, \ee
 and  $\bar{\varphi}$ is the bosonization of $\overline{U(1)}_R$ current $\bar{k}^3$,
 \be {Q\over2}\bar{\p} \bar{\varphi}=i\bar{k}^3={Q\over 2}(\bar{\p}\psi+\cos\theta\bar{\p}\phi).\ee
By definition, $\mu$ is left-moving, and $\bar{\varphi}$ is right-moving.
We define the zero mode momentum and charge by 
\be p=\oint{\p \mu \over 2\pi  },\quad \bq=Q\oint{1\over  2\pi} \left(\bp \bar{\varphi}+{2\lambda \p\mu\over Q }\right).\ee
Here  the contour is around the closed string worldsheet at fixed time. Note that the definition of $\bar q$ above is chosen in such a way that it matches the canonical momentum associated to a shift $\psi \rightarrow \psi + \epsilon$ in (\ref{TsTac}) and, thus, the quantity denoted by $\bar q$ in the supergravity analysis of the previous section. Because $\psi$ is a compact direction, $\bar q$ is naturally integrally quantized.
Letting $0 \le \sigma \le 2\pi$ denote a parameter along such a contour we find the twisted  boundary conditions
\bea \label{pd} \hat{\gamma}(\sigma+2\pi)&=& \hat{\gamma}(\sigma)-2\pi{\lambda\over Q}  (\bq-{\lambda}p),  \nn\\
\hat{\psi}(\sigma+2\pi)&=& \hat{\psi}(\sigma)+4\pi{\lambda\over Q}  p\,.
\label{bc}\eea

To conclude, string theory on the TsT background is the same as string theory on $AdS_3\times S^3$ with the twisted boundary condition (\ref{bc}). Beyond the sigma model/supergravity expansion, we will simply $define$ string theory on the TsT background as the theory obtained
by twisting the boundary condition on $AdS_3\times S^3$ according to (\ref{bc}).

Note that for  string with charge $\bar q$
the $\sigma=2\pi$ end of the string is shifted along the null direction $\hat{\gamma}$ 
by {$2\pi \lambda \bar{q}/Q$} from the $\sigma=0$ end. These strings resemble the dipole strings of \cite{Ganor:2002ju}, and have a nonlocal character. Waves along the string are continuous across $\sigma=0$ and can jump instantaneously in the $\hat{\gamma}$ direction. Hence in these variables the physics is nonlocal.

The great advantage of working with the hatted variables is that their local behavior is given by the $AdS_3 \times S^3$ sigma model. Therefore, we can use the standard OPEs from that case to construct vertex operators and solve for the spectrum of the theory. 

\subsection{Spectral flow}

We will now show that the twisted boundary conditions (\ref{bc}) define a  spectral flow with respect to the charges $(p,\bar q)$.  The magnitude of the spectral flow is dependent on the sector of the Hilbert spaces labelled by
$(p,\bar q)$.  Nevertheless the standard spectral flow formulae may be applied to obtain the deformed spectrum. 

\subsubsection{Vertex operators}

In this section we describe the vertex operators $V_{p,\bar q}$ creating states of definite $(p, \bar q)$. 
In the free field approximation to the $AdS_3 \times S^3$ model, we have the leading OPEs\bea \mu(z)\tg(w)&\sim&-\ln (z-w),\label{ope2}\\  \rho(z,\bz)\rho(w, \b{w})&\sim&- {1\over 2(Q-2)}\ln((z-w)(\bz-\bar{w} )),\label{operho} \\
\bar{\varphi}(\bar{z}) \bar{\varphi}(\bar{w})&\sim&- {2\over Q}\ln (\bar{z}-\bar{w}).
\label{ope}\eea
Corrections to the free field approximation produced by insertions of $\beta \bar \beta e^{-2\rho}$ will not produce any phases in OPEs.
By definitions $V_{p,\bar q}$ must satisfy:\footnote{The contour integrals here may be taken around the operator insertion, with a time ordering implied. 
Alternately we could take the contour around a closed string at fixed time, and replace these expressions with commutators.}
\bea
\oint{\p \mu \over 2\pi  } V_{p, \bar q} &=& p V_{p, \bar q}\,,\nonumber\\
Q\oint{1\over  2\pi} \left(\bp \bar{\varphi}+{2\lambda \p\mu\over Q }\right)  V_{p, \bar q} &=& \bq\, V_{p, \bar q}\,.
\eea
Moreover the monodromy condition (\ref{pd}) for physical states requires the log terms in the OPE 
\bea \hat \gamma(z,\bar z) V_{p, \bar q} (w,\bw)&\sim& i{\lambda\over Q} (\bq-{\lambda}p)\ln (z-w) \, V_{p, \bar q}(w,\bw), \nonumber\\
 \hat{\psi}(z,\bar z) V_{p, \bar q}(w,\bw)&\sim& 2i{\lambda\over Q}  p  \ln (\bz-\bw) V_{p, \bar q}(w,\bw) .\eea  Using the OPEs (\ref{ope2})-(\ref{ope}), it is easy to show that the following vertex operators have the right properties:
\be V_{p,\bq}=\hat V_{p, \bar q}\, e^{ip \tg} e^{i({\bq\over2} - \lambda p)\bar{\varphi}} e^{-i{{\lambda\over Q }(\bq-{\lambda}p)}\mu}\label{vertex},\ee
where $\hat V_{p, \bar q}$ has no  $U(1)$ charges. Since  $\g$ is non-compact and $\varphi\sim \varphi+4\pi$ \be \label{qc} p\in\mathbb{R},\quad \bar{q}\in \mathbb Z\,.\ee

The vertex operator (\ref{vertex}) is just the undeformed operator $V_{(0) p,\bq}$ in the $AdS_3 \times S^3$ model  dressed by the spectral flow operator $\mathcal{U}$
\be V_{p,\bq}=V_{(0) p,\bq}\, \mathcal{U},\label{spectral}\ee
where
 \bea \label{spv}
\mathcal{U}&=&e^{-i{\lambda p}\bar{\varphi}}\cdot e^{-i{{\lambda\over Q }(\bq-{\lambda}p)}\mu}.\nn
\eea
acts on both the left and the right. 
Using the free field approximation (\ref{ope2}) and (\ref{ope}), it is easy to see that $\mathcal{U}$ does create the needed branch cuts producing (\ref{bc}).

We need to show that the deformed vertex operators are mutually local. 
It is easily seen that the OPE between two operators $V_{p,\bq}({z})$ and  $V_{p',\bq'}(w)$ acquires the $\lambda$-dependent prefactor 
  \be  \bigl( (\bar{z}-\bar{w})(z-w)\bigr)^{-{\lambda\over Q}((\bq-{\lambda}p) p'+(\bq'-{\lambda}p') p) } \ee
 which has no branch cuts. 
Hence, the deformation does not disturb mutual locality of the vertex operators. The spectral flow can be also viewed as a $SO(2,1)$ rotation on the Narain lattice, rotating
 $\tg,~\hat{\psi}$ to $\gamma,~\psi$.

If $V_{p,\bar q}$ is the Fourier transform with respect to $\hat \gamma$ of a boundary primary, before the deformation it contains a factor  $\Phi_{(0) h}(p,\bar{x};z,\bz)$. After the deformation, it becomes
\bea
\Phi_h(p,\bar{x};z,\bz)&=&\Phi_{(0)h}(p,\bar{x};z,\bz)\mathcal{U}.
\eea

\subsubsection{The massive spectrum}

Given that all we are doing is a spectral flow transformation of the undeformed $AdS_3 \times S^3$ model, we can easily find the worldsheet weights of our new operators. Including the renormalization of $Q$ due to quantum effects shifts $Q \rightarrow Q-2$ and the worldsheet stress tensor is
\bea T_{WS}(z)&\equiv&-{1\over Q-2}((j^3)^2-j^+j^--(k^3)^2+k^+k^-),
\eea
where $j^a$ are $SL(2,\mathbb{R})$ currents and $k^a$ are $SU(2)$ currents.

According to the spectral flow (\ref{spectral}), the deformed Virasoro and current algebra modes on the circle are given by
\bea
L_m&=&L_{(0)m}-{\lambda\over Q-2 } (\bq-{\lambda}p) j^-_{(0)m}, \label{strss2}\\
\bar{L}_m&=&\bar{L}_{(0)m}+{1\over Q-2}( ({\lambda p})^2 \delta_{m,0}-2{\lambda p }\bar{k}^3_{(0)m}),\\ \bar{k}^3_m&=&\bar{k}^{3}_{(0)m}-{\lambda p}\delta_{m,0},
\eea
where on the right hand side are the undeformed modes built from the periodic $AdS_3\times S^3$ fields.  Using the relation 
\bea j^-_{(0)0} &\equiv&-\oint{j_{(0)}^-\over 2\pi}=p,\quad \bar{k}_{(0)0}^3\equiv\oint{i\bar{k}_{(0)}^3\over 2\pi}={\bq\over2},\eea  
the on-shell condition for the deformed state in terms of the undeformed charges becomes 
\bea
 L_0={-h(h-1)+J(J+1)\over Q-2}+{(\lambda p)^2-\lambda p \bq\over Q-2}+N-a&=&0,\label{LL} \\
 {\bar L}_0={-\bar{h}(\bar{h}-1)+\bar{J}(\bar{J}+1)\over Q-2} +{(\lambda p)^2-\lambda p\bq\over Q-2}+\bar{N}-\bar{a}&=&0, \label{LRR}
\eea
where $h$ and $\bar{h}$ are the $SL(2,\mathbb{R})$ weights, 
 $J$ and $\bar{J}$ are the $SU(2)$ weights, $N$ and $\bar N$ are oscillator numbers and $a$ and $\bar a$ are contributions from ghosts and the compact $M^4$.
Note that the level matching condition is automatically preserved by the spectral flow. However in order to preserve the $L_0=0$ and $\bar{L}_0=0$  condition the value of $h$ and $\bar{h}$ must be adjusted according to (\ref{LL}) and  (\ref{LRR}).

\section{Massive string vs supergravity modes}
From (\ref{LL}) and (\ref{LRR}), we can read the on-shell condition for the primary field, and reproduce the supergravity result. Here we will not distinguish $Q$ and $Q-2$ because we are in the classical limit.
For the highest weight representation\bea
{-\bar{h}(\bar{h}-1)+{\bq\over4}(\bq+2)}-\lambda p \bq +\lambda^2p^2+Q(\bar{N}-\bar{a})&=&0\, .\label{LR}
\eea
Therefore the conformal weights will be modified to
\be
\bar{h}=\half(1+\sqrt{1+Qm^2}),\label{hstring}
\ee
where $m$ is the mass in three dimensions, \be Qm^2=4Q(\bar{N}-\bar{a})+{\bq(\bq+2)}-4\lambda p\bar{q}+4\lambda^2p^2.\ee
The weights (\ref{hstring}) and (\ref{hlsugra}) are exactly the same if we identify $M^2=4(\bar{N}-\bar{a}).$

\section{Boundary modes}
\subsection{Right-movers}
\subsubsection{Gravitons}For the right-moving boundary gravitons, $p=0,$ therefore the on-shell condition is unchanged, with spacetime conformal weights $(h,\bar{h})=(0,2)$. Boundary  gravitons are created by nontrivial diffeomorphisms. The weight $(1,1)$ vertex operator for  a general infinitesimal diffeomorphism $\zeta^\mu$ and gauge transformation $\Lambda_\mu$ is, to leading order in the $\alpha'$ sigma model expansion,
\be V_{\zeta,\Lambda} =[\mathcal{L}_{\zeta}(g_{\mu\nu}+B_{\mu\nu}) +\p_\mu \Lambda_\nu-\p_\nu\Lambda_\mu]\p X^\mu\bp X^\nu. \ee
For the nontrivial right-moving diffeomorphisms parameterized by
$e^{i \bar{p} \bg} $, the diffeomorphisms are \be\zeta^\mu\p_\mu=e^{i\bar{p}\bg}\left(\p_{\bg}-{i\bar{p}\over2}\p_
\rho+{\bar{p}^2\over2}e^{-2\rho}\p_\g\right),\quad \Lambda_\mu=0.\ee 
These vertex operators carry neither left momentum nor right charge and are unaffected by the  deformation. Hence the tower of right-moving massless gravitons remains intact.
The  right-moving boundary stress tensor is
\bea
\bar{T}_B(\bar{x})&=&\int d\bar{p} e^{-i\bar{p}\bx}\bar{T}_B(\bar{p}),
\eea
where\be \bar{T}_B(\bar{p})=-i\int {d^2z\over \pi}  V_{\zeta,\Lambda} .\ee
According to \cite{de Boer:1998pp}, the above definition is equivalent to that of \cite{Kutasov:1999xu}
\be \bar{T}_B(\bar{p})= \frac{Q}{2\pi}\oint (e^{2\rho}\bp\tg-i\bar{p}\bp \rho)e^{i\bar{p}\bg}.\ee
One can check using the OPEs (\ref{ope2}) and (\ref{operho}) that it indeed generates the Virasoro algebra
\be[\bar{T}_B(\bar{p}),\bar{T}_B(\bar{p}')]=-i(\bar{p}-\bar{p}')\bar{T}_B(\bar{p}+\bar{p}')+{\bar{p}'^2\bar{p}-\bar{p}'\bar{p}^2\over4}\bar{I}\label{bTp}\ee
with the central term \be \bar{I}=\frac{ Q}{2\pi} \oint\bp \bg\,e^{i(\bar{p}+\bar{p}')\bg}\label{cR},\ee
independent of $\lambda$. As was calculated in \cite{Giveon:2001up} and \cite{Troost:2011ud}, the central charge of the dual CFT in the classical limit is \be c_R={6\langle\,\bar{I}\,\rangle\over\delta (\bar{p}+\bar{p}')}=6Q^2.\ee
The two point functions imply $\bar{T}_B(\bar{p})$ has $(h,\bar{h})=(0,2)$.

This suggests that the right-moving conformal symmetry of the boundary theory is unaffected by the deformation. 
\subsubsection{Gluons}
In the undeformed theory, boundary $\overline{SU(2)}_R$ gluons are created by the vertex operators
\be V^3({\bar p}) =  \bar k^3\p e^{i \bar p \bar \gamma},   ~~~~~ V^\pm({\bar p}) =\bar k^\pm\p e^{i \bar p \bar \gamma}. \ee
To get the above expression, note that in 6 dimensions the boundary  gluons are a combination of diffeomorphism of the form $\zeta^\mu\p_\mu=\bar{k}^{a}e^{i \bar{ p} \bg}$ and a corresponding gauge transformation $\Lambda^a$, for example $ \Lambda^3_\mu dx^\mu={Q\over4}e^{i \bar{p}\bg} d\psi$.
The associated nontrivial  gauge transformations are generated by the contour integrals
\be G^3({\bar p}) = \frac{1}{2\pi} \oint \bar k^3e^{i \bar p \bar \gamma},   ~~~~~ G^\pm({\bar p}) =\frac{1}{2\pi} \oint \bar k^\pm e^{i \bar p \bar \gamma}, \ee
which obey a $\overline{SU(2)}_R$ Kac-Moody algebra at level
$Q^2$.
The $\overline{U(1)}_R$ subgroup generated by $G^3$ is unaffected by the deformation and
the associated Kac-Moody tower of photons remains intact. However $G^\pm$ is deformed to
\be G^{\pm,\lambda}(\bar{p}) =\frac{1}{2\pi} \oint \bar k^\pm e^{i \bar p \bar \gamma\mp 2i{\lambda\over Q}\mu}. \ee
Note that the dressing  $e^{\mp 2i{\lambda\over Q}\mu}$ is dimension zero.
However the OPE of the current in the integrand with a massive vertex operator is not chiral. For an operator of left momentum $p$, there is an extra factor of $((z-w)(\bz-\bar w))^{\pm2{\lambda\over Q} 
p}$.  Hence the on-shell single-string states are not in a representation of the global
 $\overline{SU(2)}_R$ algebra.  This raises the possibility that the operators $V^\pm(\bar p)$ might be eliminated for some choice of boundary conditions for nonzero $\lambda$.

\subsection{Left-movers}

\subsubsection{Gluons}

In the undeformed theory, $SU(2)_L$ boundary gluons are created by the vertex operators
\be V^a({ p}) =  k^a\bp e^{i  p \tg}. \ee 
The associated nontrivial  gauge transformations are generated by the contour integrals
\be G^a({ p}) = \frac{1}{2\pi}  \oint k^a e^{i  p   \tg},   \ee
which obey a $SU(2)_L$ Kac-Moody algebra at level
$k_L= Q^2$.  In the deformed theory, the expressions obtained from these by spectral flow are
\be V^a({ p}) = k^a\bar \p (e^{i  p \tg-i{\lambda} p(\bar \varphi -{\lambda \over Q}\mu)}),\ee
\be G^a({ p}) = \frac{1}{2\pi} \oint k^a e^{i  p   \tg-i{\lambda }p(\bar \varphi -{\lambda\over Q} \mu)}. \ee
For nonzero $p$ the vertex operator is no longer dimension $(1,1)$ and the generators do not form an algebra. Hence $V^a$ does not create a physical state and physical states are not in representations of $G^a(p)$, except for $p=0$. $G^a(0)= \frac{1}{2\pi} \oint k^a$ generates the unbroken global $SU(2)_L$ isometry. An onshell dimension $(1,1)$ vertex operator could be obtained by adding radial dependence and dressing with a factor of $\Phi_h$.  But this will correspond to gauge transformations which presumably either fall off too fast and become trivial or blow up and violate asymptotic boundary conditions.

One might conclude from this that left-moving boundary gluons are not part of the physical spectrum. This may ultimately prove to be the case, but one cannot immediately draw that conclusion from the preceding. The boundary gluons are linearized pure gauge excitations of the $SU(2)_L$ gauge field obeying
\be F^a=0, \ee
which are simply
\be \label{glue}A^a(\gamma)=d \epsilon^a(\gamma).\ee
The equation or its solution do not involve the metric. Hence the linearized solutions are still there when we deform the theory.

So where is the vertex operator which creates these modes? The problem is that so far we have worked in the covariant worldsheet formalism, which is in this context over-gauge fixed. The gauge symmetry is limited to gauge parameters obeying
\be \nabla^2 \epsilon^a=\lb\nabla^2_{AdS}+\frac{4\lambda^2}{Q}\p_\g^2\rb\epsilon^a =0.\ee
For $\lambda\neq 0$, $\epsilon^a=\epsilon^a(\gamma)$ is not a solution.  To remedy this problem, we may use the worldsheet BRST formalism which is spacetime gauge covariant.
Then the ghost number $(1,1)$ vertex operator creating  an arbitrary gauge mode can be written
\be \label{brs}V(\epsilon)= \{Q_{BRST}, c\epsilon_ak^a \}.\ee
While this is BRST exact, it will not in general decouple in scattering amplitudes if $\epsilon$ does not vanish at infinity.  For $\epsilon^a=\epsilon^a(\gamma)$, this vertex operator creates the mode described in gravity by (\ref{glue}).
Whether or not we include such objects is ultimately a matter of boundary conditions and outside the scope of the present work.

\subsubsection{Gravitons}
The fate of the left-moving gravitons is similar.  These are generated by  nontrivial diffeomorphisms $\zeta^{(p)}=e^{ip\gamma}(\p_\gamma-{ip\over2}\p_\rho)$ for any $\lambda$. Their wavefunctions were explicitly computed in \cite{Guica:2010sw} where they are referred to as T modes.   For nonzero $p$ spectral flow of the undeformed covariant gauge vertex operator gives a new operator which is not dimension $(1,1)$, and the massive string states are not in representations of the spectral-flowed left Virasoro. However there is a vertex operator similar to  (\ref{brs}) which creates the worldsheet version of the supergravity T mode:
\be V(\epsilon)= \{Q_{BRST}, c\zeta^{(p)}_\mu \p X^\mu \}.\ee
The zero mode generator on the worldsheet $\frac{Q}{2\pi} \oint e^{2\rho}\p \bar \gamma $ remains well behaved and generates the unbroken left translations ${U(1)_L}$.

\subsection{Crossover modes}
In the undeformed theory, left-moving worldsheet currents lift to left-moving boundary CFT currents, and right-moving worldsheet currents lift to right-moving boundary CFT currents.
In this section we will present an alternate ``crossover" possibility which is natural in the deformed theory: the zero modes of broken right-moving currents lift to currents in the boundary CFT which transform under the left-moving spacetime Virasoro. This phenomenon has been indicated in a variety of contexts \cite{Strominger:1998yg,Hartman:2008dq, Compere:2008cv,STD}. 

\subsubsection{Left translations $\rightarrow$ right Kac-Moody}
Consider the ``crossover" vector fields
\be e^{i\bar{ p} \bg}\p_\gamma,\ee which generate general diffeos of the form
$\gamma \to \g+f(\bg)$.
The associated worldsheet currents are
\be \xi_C(\bar {p})=Qe^{i\bar {p} \bg} e^{2\rho} \p\bg = e^{i\bar {p} \bg}\beta.\ee
The zero mode here $\xi_C(0)=\beta $ generates global $U(1)_L$ left translation, but we are dressing with the right momentum.
The linearized metric fluctuation is created by the vertex operator
\be \label{wc} W_C(\bar p)=\bar \p \xi_C(\bar p).\ee
The generators \be \bar G_C(\bar p)=\frac{1}{2\pi}  \oint \xi_C(\bar p) \ee
transform canonically under the unbroken right Virasoro as
\be [\bar{T}_B(\bar{p}),\bar G_C(\bar{p}')]=i\bar{p}'\bar G_C(\bar{p}+\bar{p}'), \ee
and obey a Kac-Moody algebra
\be  [\bar G_C(\bar{p}),\bar G_C(\bar{p}')]=0,\ee
with vanishing level. The level vanishes because  $\beta$ is null -- we will see other cases below where it is nonzero. 

Since these operators carry neither left momentum nor $\overline{U(1)}_R$ charge ($p$ or $\bar q$)
they are unaffected by the deformation.  In the undeformed theory, with Brown-Henneaux boundary conditions,  they would ordinarily be excluded because their action on a left-moving graviton creates a mode that grows at infinity and violates the boundary condition. Put another way, the algebra consisting of a left and a right-moving Virasoro plus a right Kac-Moody whose zero mode is the same as the left Virasoro zero mode does not close.

In the deformed theory there may be other natural options, for example boundary conditions which eliminate the broken left Virasoro, and keep the crossover Kac-Moody. Boundary conditions of this general type are discussed in slightly different contexts in \cite{Strominger:1998yg,Hartman:2008dq, Compere:2008cv}. In this way it might be possible to define a string theory in which the vertex operators $W_C(\bar p)$ in (\ref{wc}) are allowed but the broken Virasoros are not. In this paper however we concentrate on the worldsheet and leave these spacetime considerations to future work.

\subsubsection{${SU(2)_L}$ rotations $\rightarrow$ right  Kac-Moody}
Similar considerations apply to the ${SU(2)_L}$ Kac-Moody of the undeformed theory. The deformation breaks it down to the global subgroup, and the global generators can be dressed with
the right momentum. The vertex operators and generators are
\be  W^a_C(\bar p)= k^a\bp e^{i\bar p \bg},\ee
\be \bar G^a_C(\bar p)=-{i\over\pi} \int d^2z W^a_C(\bar{p}) . \ee
These form an $SU(2)_L$ crossover Kac-Moody 
\be[\bar G^a_C(\bar p),\bar G^b_C(\bar p')]=if^{abc}\bar G^c_C(\bar{ p}+\bar{p}')+(\bar{p}-\bar{p}')k^{ab} {\bar{I}\over2}, \ee
where for large $Q$ the central term is  \be \bar{I}=
{Q\over \pi\bar{p}}\int d^2z \bp e^{i\bar{p}\bg}\p e^{i\bar{p}'\bg}={Q\over2\pi}\oint\bp\bg\delta(\bar{p}+\bar{p}')=Q^2\delta(\bar{p}+\bar{p}').\ee
Similar to the $U(1)_L$ crossover, the ${SU(2)}_L$ crossover  transforms under the unbroken right-moving Virasoro  as \be [\bar{T}_B(\bar{p}),\bar G^a_C(\bar{p}')]=i\bar{p}'\bar G^a_C(\bar{p}+\bar{p}'). \ee

\section{Black strings}

If we heat up the NS1-NS5 string at left and right temperatures $T_L,~T_R$, the near horizon geometry becomes 
\bea
{4ds^2\over Q}&=&-{r^2-(2\pi^2T_L T_R)^2\over \pi^2 T_L^2}(dt^+)^2+ 4\pi^2 T_L^2\left( dt^-+{rdt^+\over2\pi^2 T_L^2}\right)^2+{dr^2\over r^2-(2\pi^2T_L T_R)^2}+4d\Omega_3^2,\nn\\
B&=&-{Q\over4} (\cos\theta d\phi \wedge d\psi+2 r dt^-\wedge dt^+ ), \label{btz}
\eea
which is still locally $AdS_3\times S^3$. The string lies in the $(t^+,t^-)$ plane. 
Applying TsT to this (T dual along ${t^-}$, shift $\psi\rightarrow\psi-2 \frac{\lambda}{Q} t^-$ and then T dual along ${t^-}$ again) gives the dual of the deformed theory at finite temperature. Transforming to the coordinates 
 \be t^-=\tau^-+{\lambda\over2}\chi,\quad \psi=\chi +2\pi^2T_L^2\lambda \tau^-.\label{Udual}\ee
the string frame geometry is 
\bea
{4ds^2\over Q}&=&-{r^2-(2\pi^2T_L T_R)^2\over \pi^2 T_L^2}(dt^+)^2+{dr^2\over r^2-(2\pi^2T_L T_R)^2}+ 4\pi^2 T_L^2\left( d\tau^-+{rdt^+\over2\pi^2 T_L^2}\right)^2\nn\\
&&+4d\Omega_3^2+4\pi T_L\alpha\left(d\tau^-+{rdt^+\over2\pi^2 T_L^2}\right) (d\chi+\cos\theta d\phi),\nn\\
B&=&-{Q\over4} \lb \cos\theta d\phi \wedge d\chi+2rd\tau^- \wedge dt^+ +2\pi T_L\alpha (d\chi+\cos\theta d\phi)\wedge\left(d\tau^-+{r dt^+\over 2\pi^2 T_L^2}\right)\rb\label{TsTBTZ},\nn\\
e^{-2\phi}&=&1+(\lambda\pi T_L)^2,\nn\\
\alpha&=&{2\lambda\pi T_L\over1+(\lambda \pi T_L)^2}.
\eea

It is easy to check that the entropy density per unit  $x={t^++t^-\over2}$ is unaffected by the TsT transformation and is given by  \be {\delta S\over \delta x}={\pi\over6}(c_LT_L+c_RT_R) ,\quad c_L=c_R=6Q^2.\ee
where we have evaluated the entropy density along the line.

The worldsheet sigma model on this warped black string can be transformed back to 
(\ref{btz}) by the field redefinition
\bea
\p\hat{\chi}&=&\p\chi+{2\over Q}\alpha j^1,\quad  2\pi T_L\bp\hat{\tau}^-=2\pi T_L\bp \tau^-+{2\over Q}\alpha \bar{k}^3,
\\
\bp\hat{\chi}&=&\bp\chi+{2\over Q}(\sqrt{1-\alpha^2}-1)\bar{k}^3,\quad 2\pi T_L\p\hat{\tau}^-=2\pi T_L\p\tau^-+{2\over Q}(\sqrt{1-\alpha^2}-1)j^1,\nn
\eea 
where
\bea
j^1={Q\over2}\left(2\pi T_L\p \tau^- +{r \p t^+\over\pi T_L}\right),\quad \bar{k}^3={Q\over2}(\bp\chi+\cos\theta\bp\phi)\eea satisfy $\bp j^1=0$ and $\p \bar{k}^3=0$.
Define the bosonization of the two $U(1)$ currents \be \p y=2{j^1\over Q},\quad \bp\bar{\varphi}=2{\bar{k}^3\over Q}. \ee   
Vertex operators in the form $V=V_0e^{i(p_y\hat {y}+{\bar{q}\over2}\hat{\bar{\varphi}})}$ will be deformed to 
\be\tilde{V}=V_0e^{{i\over\sqrt{1-\alpha^2}}\lb(p_y-\alpha{\bar{q}\over2})\hat{y}+({\bar{q}\over2}-\alpha p_y)\hat{\bar{\varphi}}\rb }.\ee
Note that the $U(1)$ from the $SL(2,\mathbb{R})$ is space-like now, so we have the OPE
\be \hat{y}(z)\hat{y}(w)\sim -{2\over Q}\ln(z-w),\quad \hat{\bar{\varphi}}(\bar{z})\hat{\bar{\varphi}}(\bar{w})\sim  -{2\over Q}\ln(\bz-\bw).\ee
Using the OPEs above, we can show that the worldsheet weights are shifted to 
\bea L_0 \rightarrow L_0+{\alpha^2 (p_y^2+{\bar{q}^2\over4})-\alpha p_y\bar{q}\over (Q-2) (1-\alpha^2)},\\
\bar{L}_0 \rightarrow \bar{L}_0+{\alpha^2 (p_y^2+{\bar{q}^2\over4})-\alpha p_y\bar{q}\over (Q-2) (1-\alpha^2)},\
\eea
where $p_y={p_{\tau^-}\over 2\pi T_L}\rightarrow p$ in the zero temperature limit $T_L\rightarrow 0$.
Again, this reproduces the supergravity result in the large $Q$ limit.

The $U(1)_L$ crossover vector fields are  \bea   \zeta^\mu\p_\mu&=&{e^{i\bar {p} t^+}\over2\pi T_L}{\p\over\p \tau^-},\\ 
\Lambda_\mu dx^\mu&=&{Q\pi T_L\over2}e^{i\bar{p}t^+} d \tau^- .
\eea
The corresponding vertex operator at large radius is
\bea
W_C(\bar p)
 &=& {Q\over2}\left(2\pi T_L\bp \hat{\tau}^- +{r \bp t^+\over\pi T_L}\right)\p e^{i\bar{p}t^+}.\eea
Note that at large $r\sim e^{2\rho},\quad\hat{\tau}^-\sim \tg,\quad t^+\sim\bg.$ The central term is \bea \bar{I}={Q\over\pi \bar{p}}\int d^2z \bp e^{i\bar{p}t^+}\p e^{i\bar{p}'t^+}=Q^2\delta(\bar{p}+\bar{p}').\eea

\section*{Acknowledgements}
We are grateful to A.~Castro, K.~Dasgupta, S.~Detournay, T.~Hartman, J.~Maldacena, A.~Maloney and B. van Rees for useful conversations. 
This work was supported in part by DOE grant DE-FG02-91ER40654. T.A. was in part supported by JSPS Postdoctoral Fellowship for Research Abroad. W.S. was also supported in part by the Harvard Society of Fellows and the National Science Foundation under Grant No. NSF PHY11-25915.
 

\begin{thebibliography}{99}

\bibitem{Anninos:2008fx}
  D.~Anninos, W.~Li, M.~Padi, W.~Song and A.~Strominger,
  ``Warped AdS(3) Black Holes,''
  JHEP {\bf 0903}, 130 (2009)
  [arXiv:0807.3040 [hep-th]].
\bibitem{Guica:2008mu}
  M.~Guica, T.~Hartman, W.~Song and A.~Strominger,
  ``The Kerr/CFT Correspondence,''
  Phys.\ Rev.\ D {\bf 80}, 124008 (2009)
  [arXiv:0809.4266 [hep-th]].
  
\bibitem{Bergman:2000cw}
  A.~Bergman and O.~J.~Ganor,
  ``Dipoles, twists and noncommutative gauge theory,''
  JHEP {\bf 0010}, 018 (2000)
  [arXiv:hep-th/0008030].

\bibitem{Dasgupta:2000ry}
  K.~Dasgupta, O.~J.~Ganor and G.~Rajesh,
  ``Vector deformations of N=4 superYang-Mills theory, pinned branes, and
  arched strings,''
  JHEP {\bf 0104}, 034 (2001)
  [arXiv:hep-th/0010072].

\bibitem{Bergman:2001rw}
  A.~Bergman, K.~Dasgupta, O.~J.~Ganor, J.~L.~Karczmarek, G.~Rajesh,
  ``Nonlocal field theories and their gravity duals,''
  Phys.\ Rev.\  {\bf D65}, 066005 (2002).
  [hep-th/0103090].

  
\bibitem{ElShowk:2011cm}
  S.~El-Showk and M.~Guica,
  ``Kerr/CFT, dipole theories and nonrelativistic CFTs,''
  arXiv:1108.6091 [hep-th].
\bibitem{Song:2011sr}
  W.~Song and A.~Strominger,
  ``Warped AdS3/Dipole-CFT Duality,''
  JHEP {\bf 1205}, 120 (2012)
  [arXiv:1109.0544 [hep-th]].

\bibitem{Son:2008ye}
  D.~T.~Son,
  ``Toward an AdS/cold atoms correspondence: A Geometric realization of the Schrodinger symmetry,''
  Phys.\ Rev.\ D {\bf 78}, 046003 (2008)
  [arXiv:0804.3972 [hep-th]].

\bibitem{Balasubramanian:2008dm}
  K.~Balasubramanian and J.~McGreevy,
  ``Gravity duals for non-relativistic CFTs,''
  Phys.\ Rev.\ Lett.\  {\bf 101}, 061601 (2008)
  [arXiv:0804.4053 [hep-th]].
\bibitem{Hartnoll:2009sz}
  S.~A.~Hartnoll,
  ``Lectures on holographic methods for condensed matter physics,''
  Class.\ Quant.\ Grav.\  {\bf 26}, 224002 (2009).
  [arXiv:0903.3246 [hep-th]].
\bibitem{Lunin:2005jy}
  O.~Lunin and J.~M.~Maldacena,
  ``Deforming field theories with U(1) x U(1) global symmetry and their gravity duals,''
  JHEP {\bf 0505}, 033 (2005)
  [hep-th/0502086].

\bibitem{Maldacena:2008wh}
  J.~Maldacena, D.~Martelli and Y.~Tachikawa,
  ``Comments on string theory backgrounds with non-relativistic conformal symmetry,''
  JHEP {\bf 0810}, 072 (2008)
  [arXiv:0807.1100 [hep-th]].

\bibitem{Detournay:2010rh} 
  S.~Detournay, D.~Israel, J.~M.~Lapan and M.~Romo,
  ``String Theory on Warped $AdS_{3}$ and Virasoro Resonances,''
  JHEP {\bf 1101}, 030 (2011)
  [arXiv:1007.2781 [hep-th]].

\bibitem{Maldacena:2000hw}
  J.~M.~Maldacena and H.~Ooguri,
  ``Strings in AdS(3) and SL(2,R) WZW model 1.: The Spectrum'
  J.\ Math.\ Phys.\  {\bf 42}, 2929 (2001)
  [hep-th/0001053].

\bibitem{Balog:1988jb} 
  J.~Balog, L.~O'Raifeartaigh, P.~Forgacs and A.~Wipf,
  ``Consistency of String Propagation on Curved Space-Times: An SU(1,1) Based Counterexample,''
  Nucl.\ Phys.\ B {\bf 325}, 225 (1989).
  
\bibitem{Bars:1990rb} 
  I.~Bars and D.~Nemeschansky,
  Nucl.\ Phys.\ B {\bf 348}, 89 (1991).
  
\bibitem{Evans:1998qu} 
  J.~M.~Evans, M.~R.~Gaberdiel and M.~J.~Perry,
  Nucl.\ Phys.\ B {\bf 535}, 152 (1998)
  [hep-th/9806024].
  
\bibitem{Bars:1995cn} 
  I.~Bars,
  In *Erice 1995, String gravity and physics at the Planck energy scale* 151-169, and South. Calif. U. Los Angeles - USC-95-HEP-B03 (95,rec.Dec.) 18 p
  [hep-th/9511187].



\bibitem{Frolov:2005dj}
  S.~Frolov,
  ``Lax pair for strings in Lunin-Maldacena background",
  JHEP {\bf 0505}, 069 (2005)
  [arXiv:hep-th/0503201].

\bibitem{Alday:2005ww}
  L.~F.~Alday, G.~Arutyunov and S.~Frolov,
  ``Green-Schwarz strings in TsT-transformed backgrounds",
  JHEP {\bf 0606}, 018 (2006)
  [arXiv:hep-th/0512253].


\bibitem{Klebanov:1999tb} 
  I.~R.~Klebanov and E.~Witten,
  ``AdS / CFT correspondence and symmetry breaking,''
  Nucl.\ Phys.\ B {\bf 556}, 89 (1999)
  [hep-th/9905104].

\bibitem{Guica:2011ia}
  M.~Guica,
  ``A Fefferman-Graham-Like Expansion for Null Warped AdS(3),''
  arXiv:1111.6978 [hep-th].

\bibitem{balt}
 B.~C.~van Rees,
  ``Correlation functions for Schr\'odinger backgrounds,''
  arXiv:1206.6507 [hep-th].

\bibitem{Strominger:1998yg}
  A.~Strominger,
  ``AdS(2) quantum gravity and string theory,''
  JHEP {\bf 9901}, 007 (1999)
  [hep-th/9809027].
  
\bibitem{Hartman:2008dq}
  T.~Hartman and A.~Strominger,
  ``Central Charge for AdS(2) Quantum Gravity,''
  JHEP {\bf 0904}, 026 (2009)
  [arXiv:0803.3621 [hep-th]].
  
\bibitem{STD}
  S.~Detournay, T.~Hartman and D.~M.~Hofman,
  to appear.

\bibitem{Guica:2010sw}
  M.~Guica, K.~Skenderis, M.~Taylor and B.~C.~van Rees,
  ``Holography for Schrodinger backgrounds,''
  JHEP {\bf 1102}, 056 (2011)
  [arXiv:1008.1991 [hep-th]].

\bibitem{Hofman:2011zj}
  D.~M.~Hofman and A.~Strominger,
  ``Chiral Scale and Conformal Invariance in 2D Quantum Field Theory,''
  Phys.\ Rev.\ Lett.\  {\bf 107}, 161601 (2011)
  [arXiv:1107.2917 [hep-th]].

 

\bibitem{Giveon:1998ns}
  A.~Giveon, D.~Kutasov, N.~Seiberg,
  ``Comments on string theory on AdS(3),"
  Adv.\ Theor.\ Math.\ Phys.\  {\bf 2}, 733-780 (1998).
  [hep-th/9806194].

\bibitem{Kutasov:1999xu}
  D.~Kutasov, N.~Seiberg,
  ``More comments on string theory on AdS(3),"
  JHEP {\bf 9904}, 008 (1999).
  [hep-th/9903219].
  
\bibitem{Maldacena:1998bw}
  J.~M.~Maldacena, A.~Strominger,
 ``AdS(3) black holes and a stringy exclusion principle,"
  JHEP {\bf 9812}, 005 (1998).
  [hep-th/9804085].

\bibitem{Ganor:2002ju}
  O.~J.~Ganor and U.~Varadarajan,
  ``Nonlocal effects on D-branes in plane wave backgrounds,"
  JHEP {\bf 0211}, 051 (2002)
  [hep-th/0210035].
  
\bibitem{de Boer:1998pp}
  J.~de Boer, H.~Ooguri, H.~Robins and J.~Tannenhauser,
  ``String theory on AdS(3),''
  JHEP {\bf 9812}, 026 (1998)
  [hep-th/9812046].
  
  \bibitem{Giveon:2001up}
  A.~Giveon and D.~Kutasov,
  ``Notes on AdS(3),''
  Nucl.\ Phys.\ B {\bf 621}, 303 (2002)
  [hep-th/0106004].

  \bibitem{Troost:2011ud}
  J.~Troost,
  ``The $AdS_3$ central charge in string theory,''
  Phys.\ Lett.\ B {\bf 705}, 260 (2011)
  [arXiv:1109.1923 [hep-th]].
  
\bibitem{Compere:2008cv} 
  G.~Compere and S.~Detournay,
  ``Semi-classical central charge in topologically massive gravity,''
  Class.\ Quant.\ Grav.\  {\bf 26}, 012001 (2009)
  [Erratum-ibid.\  {\bf 26}, 139801 (2009)]
  [arXiv:0808.1911 [hep-th]].


\end{thebibliography}
\end{document}